\documentclass[12pt]{article}
\usepackage[reqno]{amsmath}
\usepackage{epsfig}
\usepackage{array}
\usepackage{float}
\usepackage{dsfont}
\usepackage{amstext}

\usepackage{a4}

\usepackage{a4wide}
\def\ra{\rightarrow}
\def\be{\begin{equation}}
\def\ee{\end{equation}}
\def\gs{\mathrel{
   \rlap{\raise 0.511ex \hbox{$>$}}{\lower 0.511ex \hbox{$\sim$}}}}
\def\ls{\mathrel{
   \rlap{\raise 0.511ex \hbox{$<$}}{\lower 0.511ex \hbox{$\sim$}}}}

\newcommand{\ba}{\begin{array}{c}}
\newcommand{\baz}{\begin{array}{cc}}
\newcommand{\bad}{\begin{array}{ccc}}
\newcommand{\bea}{\begin{equation} \begin{array}{c}}
\newcommand{\eea}{ \end{array} \end{equation}}
\newcommand{\ea}{\end{array}}
\newcommand{\D}{\displaystyle}
\newcommand{\dms}{\mbox{$\Delta m^2_{\odot}$}}
\newcommand{\dma}{\mbox{$\Delta m^2_{\rm A}$}}
\newcommand{\meff}{\mbox{$\langle m \rangle$}}

\begin{document}

\title{\vspace{-.1cm}
\hfill {\small arXiv: 0905.2126 [hep-ph]}
\vskip 0.4cm
\bf 
Scaling in the Neutrino Mass Matrix, 
$\mu$--$\tau$ Symmetry and the See-Saw Mechanism}
\author{
Anjan S.~Joshipura$^{a}$\thanks{email: 
\tt anjan@prl.res.in}\mbox{ },~~
Werner Rodejohann$^{b}$\thanks{email: 
\tt werner.rodejohann@mpi-hd.mpg.de} 
\\\\
{\normalsize \it $^a$Physical Research Laboratory,}\\
{\normalsize \it Navrangpura, Ahmedabad 380 009, Gujarat, India}\\ \\
{\normalsize \it$^b$Max--Planck--Institut f\"ur Kernphysik,}\\
{\normalsize \it  Postfach 103980, D--69029 Heidelberg, Germany} 
} 
\date{}
\maketitle
\thispagestyle{empty}
\vspace{-0.8cm}
\begin{abstract}
\noindent  

The scaling hypothesis postulates proportionality of two 
columns of the Majorana neutrino mass matrix in the flavor basis. 
This Ansatz was shown to lead to an inverted hierarchy and 
$U_{e3}=0$. We discuss theoretical and phenomenological 
properties of this hypothesis. We show that (i) 
the neutrino mass matrix with scaling follows as a consequence of
a generalized $\mu$--$\tau$ symmetry imposed on the 
type-I see-saw model; (ii) 
there exists a unique texture for the Dirac mass matrix 
$m_D$ which leads to scaling for arbitrary Majorana 
matrix $M_R$ in the context of the type-I see-saw mechanism; (iii) 
unlike in the $\mu$--$\tau$ symmetric case, a simple model 
with two right-handed neutrinos and scaling can lead 
to successful leptogenesis both with and without the 
inclusion of flavor effects.

\end{abstract}

\newpage
\section{\label{sec:intro}Introduction}

There have been many approaches to explain the peculiar (both 
measured and possible) features of neutrino masses and lepton mixing, 
for recent reviews see e.g.~\cite{revs,WR}. Very often, the see-saw 
mechanism \cite{I}, and mostly the type I variant, is used 
to explain the smallness of neutrino masses. As additional motivation,
the see-saw mechanism contains all necessary ingredients to produce
the baryon asymmetry of the Universe via leptogenesis
\cite{lepto}. 

Unfortunately, the number of parameters in general
see-saw scenarios exceeds the measurable ones, and full 
reconstruction of the see-saw parameters is at least very difficult 
\cite{diffi}. Moreover, irrespective of their origin, it seems very 
unlikely that (ranges of) all the elements of the neutrino mass 
matrix can be determined purely from experiments. This has led to
postulates of various Ans\"atze or symmetries 
for the neutrino mass matrix in order to have more predictivity. 
Texture zeros or $\mu$--$\tau$ symmetry are popular examples for such 
strategies. 
Another Ansatz, proposed in Refs.~\cite{scal1,scal2}, is called 
scaling\footnote{Similar mass 
matrices with this form have been obtained in specific models in 
\cite{scal0,scal_others}.}. 
In this letter we wish to show that the scaling hypothesis is 
deeply connected to a generalized version of the 
$\mu$--$\tau$ symmetry and can follow from such symmetry. 
We also study yet unexplored  implications of the scaling 
hypothesis on the see-saw structure and at the 
phenomenological level on leptogenesis and lepton 
flavor violation.

We will first, in Section \ref{sec:md}, 
discuss properties of the Dirac mass matrix $m_D$ in general  
see-saw frameworks incorporating scaling. We show that 
the scaling hypothesis uniquely determines its structure. 
Moreover, scaling is shown to follow from a generalized $\mu$--$\tau$
invariance applied to the type-I see-saw. 
As the possibility of leptogenesis 
is one important consequence of the see-saw mechanism, we will 
investigate leptogenesis in the context of the scaling hypothesis 
in Section \ref{sec:yb}. 
First we study a simple two right-handed neutrino case, followed 
by a three right-handed neutrino case generated by a 
$Z_{2L} \times Z_{2R}$ symmetry. In the case of two right-handed
neutrinos we show that the baryon asymmetry for unflavored 
leptogenesis is proportional to the solar neutrino 
mass-squared difference. The three neutrino model 
is shown to lead to the same result. 
Details of the related case of $\mu$--$\tau$ symmetric see-saw with 
two heavy neutrinos are delegated to the Appendix, where we also 
summarize relevant formulae for flavored and unflavored 
leptogenesis. We conclude and summarize in 
Section \ref{sec:concl}. In the following, to set the stage, 
we will first summarize the properties and predictions of scaling.

\section{\label{sec:scal}Neutrino Mixing, Scaling and 
generalized $\mu$--$\tau$ Symmetry}
In the charged lepton basis, neutrino mass and lepton mixing originates 
at low energies from the following  neutrino mass matrix 
appearing in the Lagrangian 
\be
{\cal L} = \frac 12 \, \overline{\nu_L^c} \, m_\nu \, \nu_L 
+ h.c. 
\ee
Diagonalization of $m_\nu$ is achieved via 
$U^\ast \, m_\nu^{\rm diag} \, U^\dagger = m_\nu$, where $U$ is the 
Pontecorvo-Maki-Nakagawa-Sakata (PMNS) matrix, whose standard 
parametrization is  
\be
U = \left(
\bad  
c_{12} \, c_{13} & s_{12} \, c_{13} & s_{13}
\, e^{- i \delta}
\\[0.2cm] 
-s_{12}  \, c_{23} - c_{12}  \, s_{23} 
\, s_{13}  \, e^{i \delta} 
& c_{12}  \, c_{23} - s_{12}  \, s_{23}  
\, s_{13}  \, e^{i \delta} 
& s_{23}  \, c_{13} 
\\[0.2cm] 
s_{12}  \, s_{23} - c_{12}  \, c_{23}  
\, s_{13}  \, 
e^{i \delta} & 
- c_{12}  \, s_{23} - s_{12}  \, c_{23}  
\, s_{13}  \, e^{i \delta} 
& c_{23}  \, c_{13} 
\ea 
\right) P \, .
\ee
Here $c_{ij} = \cos \theta_{ij}$, $s_{ij} = \sin \theta_{ij}$ 
and $P = {\rm diag}(1, \, e^{i\alpha} , \, e^{i\beta})$ contains 
the Majorana phases. 

Because neutrino data implies that $\theta_{23}$ is (close to) maximal
and $\theta_{13}$ (close to) zero, many works have been devoted to 
$\mu$--$\tau$ symmetry \cite{rabi_mutau,mt,mt0}. It is defined as an 
interchange symmetry $\nu_{\mu L} \leftrightarrow \nu_{\tau L}$ in the
diagonal charged lepton basis. The generator of this $Z_2$ is 
\be \label{eq:Smt}
S_{\rm \mu\tau} = \left( 
\bad 
1 & 0 & 0 \\ 
0 & 0 & 1 \\
0 & 1 & 0 
\ea
\right)
\ee
and the resulting mass matrix is obtained from the requirement 
$S^{-1}_{\rm \mu\tau} \, m_\nu \, S_{\rm \mu\tau} = m_\nu$ and reads 
(unless stated otherwise, all entries in the mass matrices 
here and in the following can be complex) 
\be \label{eq:mnumt}
m_\nu = \left( 
\bad 
a & b & b \\ 
\cdot & d & e \\
\cdot & \cdot & d
\ea
\right). 
\ee
One eigenvector of this matrix is $(0,-1,1)^T$. If the eigenvalue 
$|d - e|$ corresponding to this eigenvector is the largest or 
the smallest one, then one predicts $\theta_{23} = \frac{\pi}{4}$ and 
$\theta_{13} = 0$ in the normal or inverted mass ordering,
respectively. 

It is possible to replace the (left) $\mu$--$\tau$ symmetry by a
more general one which leads to the prediction $\theta_{13}=0$ 
but allows arbitrary $\theta_{23}$. This generalization 
discussed in \cite{mt0} amounts to imposing the 
following $Z_2$ symmetry:
\be \label{eq:S}
S(\theta) =  
\left( 
\bad 
1 & 0 & 0 \\ 
0 & \cos 2 \theta & \sin 2 \theta \\
0 & \sin 2 \theta & -\cos 2 \theta 
\ea
\right) , 
\ee
It follows that $S^2=1$,
%$S^2 = \mathbbm{1}$, 
independent of $\theta$. Therefore $S$ defines a $Z_2$ symmetry. 
Invariance of $m_\nu$ under this $Z_2$ leads to the 
prediction $\theta_{23} = \theta$ and $\theta_{13} = 0$. Obviously, 
with $\theta = \frac{\pi}{4}$ the generator $S$ coincides 
with the one in Eq.~(\ref{eq:Smt}): 
\be \label{eq:mt}
S(\theta\rightarrow \pi/4) =  
\left( 
\bad 
1 & 0 & 0 \\ 
0 & 0 & 1 \\
0 & 1 & 0 
\ea
\right) = S_{\mu \tau}.  
\ee
The $Z_2$ symmetry defined by Eq.~(\ref{eq:S}) can therefore be 
considered as generalized $\mu$--$\tau$ symmetry \cite{mt0}. 
Note that in order to identify $\theta_{23} = \theta$ and 
$\theta_{13} = 0$ in the normal or inverted mass ordering, 
it is necessary that the eigenvalue of the 
corresponding eigenvector $(0,\tan\theta,1)^T$ 
is the largest or smallest one.\\ 

The scaling Ansatz, 
first proposed in \cite{scal1} and further analyzed in \cite{scal2}, 
is more predictive than the 
(generalized) $\mu$--$\tau$ symmetry. It postulates a 
particular structure of the 
low energy neutrino mass matrix: 
\be \label{eq:mnu}
m_\nu = 
\left( 
\bad 
a & b & b/c \\
\cdot & d & d/c \\
\cdot & \cdot & d/c^2
\ea 
\right) .
\ee
This form originates from the requirement that the ratio 
$(m_\nu)_{\alpha \mu}/(m_\nu)_{\alpha \tau}$ equals the 
``scaling factor'' $c$ for all $\alpha = e, \, \mu, \, \tau$. 
We summarize here some properties of scaling: 
\begin{itemize}
\item the mass matrix has rank 2, i.e., one vanishing mass 
eigenvalue. The eigenvector for this eigenvalue is 
$(0, \, -1, \, c)$. Hence, the matrix is only compatible with an
inverted hierarchy and predicts that 
$U_{e3} = 0$ and $\tan^2 \theta_{23} = 1/c^2$. 
There is no CP violation in oscillation experiments and 
atmospheric neutrino mixing is in general non-maximal. 
Solar neutrino mixing is naturally large, but not specified; 

\item in contrast to the flavor symmetry $L_e - L_\mu - L_\tau$  
\cite{lelmlt}, which has frequently 
been used to obtain an inverted hierarchy,  
scaling requires no breaking of the symmetry 
in order to generate successful phenomenology. 
Recall that $L_e - L_\mu - L_\tau$ predicts two degenerate eigenvalues, 
maximal $\theta_{12}$, and breaking terms which are at least 30 \% the 
magnitude of the terms allowed by the symmetry; 
\item there can be one physical phase, which is a 
low energy Majorana phase. It would appear 
in the effective mass governing neutrino-less double beta decay: 
$\meff \simeq \sqrt{\dma} \, \sqrt{1 - \sin^2 2 \theta_{12} 
\, \sin^2 \alpha}$. The scaling Ansatz is therefore fully reconstructible; 
\item interestingly, if not more than one Higgs doublet is present, 
the beta-functions of both $m_3$ and 
$\theta_{13}$ are proportional to $m_3$. Thus, the prediction 
$m_3 = U_{e3} = 0$ is not modified by radiative corrections, as has
been noted in the first paper of Ref.~\cite{scal0}.
\end{itemize}

We stress here that scaling leads to an inverted hierarchy, 
vanishing $\theta_{13}$ and arbitrary $\theta_{23}$. However, in 
contrast to generalized $\mu$--$\tau$ symmetry, there is no ambiguity 
with regards to the identification of the mass eigenvalues, 
the angles $\theta_{13} = 0$ and $\theta_{23}$ always belong to the 
vanishing mass. 

All features of scaling are attractive and easily testable, but rely 
on the low energy part of lepton physics. 
Let us note here that the analysis that we will present applies to 
the case of a real 
and diagonal charged lepton mass matrix, i.e., in the natural limit of
negligible charged lepton corrections to the mixing angles. A model
based on the flavor symmetry $D_4 \times Z_2$ in which the charged
lepton mass matrix is diagonal and $m_\nu$ obeys scaling has been
constructed in Ref.~\cite{scal1}. Nevertheless, in this paper we 
will not consider any specific model, but will continue with 
a discussion of scaling in the charged lepton basis and in the 
type I see-saw mechanism. 

We will show in the next Section that the 
scaling is a consequence of the generalized $\mu$--$\tau$ 
symmetry imposed on the type-I see-saw model.

\section{\label{sec:md}Scaling and the See-Saw Mechanism}
It was noted in \cite{scal1} that the scaling form for 
$m_\nu$ follows in the type-I see-saw model 
for an arbitrary $M_R$ if $m_D$ has the form:
\be \label{eq:md}
m_D = 
\left(
\bad 
a_1 & b & b/c \\ 
a_2 & d & d/c \\ 
a_3 & e & e/c 
\ea
\right) .
\ee
We will prove next a theorem which states 
that this is {\it the only} allowed $m_D$ if $m_\nu$ is
to have the form Eq.~(\ref{eq:mnu}). 

\subsection{\label{sec:theo} A Theorem for $m_D$}

We claim that the most general $m_D$ which can reproduce the 
scaling form for $m_\nu$ is given by Eq.~(\ref{eq:md}) 
if $m_\nu$ is to posses two non-zero eigenvalues. 
To see this, we first note that scaling implies that 
the low energy mass matrix has an eigenvector of the form 
\be \label{eq:psi}
m_\nu \, | \psi \rangle = 0 \mbox{ where } | \psi \rangle 
= \left( 
\ba
0 \\ -1 \\ c 
\ea
\right) .
\ee
The fact that one mass is zero implies that det$(m_\nu) = 0$ and 
therefore 
\be
{\rm det}(m_D) = 0 \,. 
\ee
Recall that $M_R$ needs to be non-singular in order to have a 
valid type I see-saw. The above relation implies
now for $m_D$ that 
\be
m_D \, | \chi \rangle = 0
\ee
for one of its eigenvectors $| \chi \rangle$. With this expression 
and the definition of $m_\nu$ it also follows that 
\be \label{eq:chi}
m_\nu \, | \chi \rangle = 0 \, .
\ee
Equations (\ref{eq:psi}) and (\ref{eq:chi}) together imply that
either $m_\nu$ possesses two massless states, or that 
$| \chi \rangle$ is proportional to $| \psi \rangle$. The first
possibility cannot generate two different scales unless there are huge
perturbations acting in $m_\nu$. Instead, one is lead to the second
possibility, which means 
\be
m_D \, | \psi \rangle = 0 \, .
\ee
With $| \psi \rangle$ defined in Eq.~(\ref{eq:psi}), 
it is straightforward to show that the most general solution of 
this equation is indeed given by Eq.~(\ref{eq:md}). 
Note that we have not made any assumptions about 
the structure of $M_R$, other than it is non-singular.

There is another way to show 
that Eq.~(\ref{eq:md}) is the implied form of
$m_D$: consider the following parametrization of the Dirac mass matrix
\cite{CI}: 
\be \label{eq:CI}
m_D = i \, \sqrt{M_R} \,  R \, \sqrt{m_\nu^{\rm diag}} 
\,  U^\dagger \, ,
\ee
where $R$ is a complex and orthogonal matrix. Here,  $M_R$ 
need not be diagonal. Inserting $m_3 = U_{e3} = 0$ in this
parametrization will show that 
$(m_D)_{12}/(m_D)_{13} = (m_D)_{22}/(m_D)_{23} 
= (m_D)_{32}/(m_D)_{33} = -\cot \theta_{23}$, 
independent of $M_R$ and $R$. Since $\cot \theta_{23} = c$, the same 
form for $m_D$ as in Eq.~(\ref{eq:md}) is implied.

\subsection{\label{sec:S2}Scaling and $Z_2$ Symmetries} 

The scaling form as given in Eq.~(\ref{eq:mnu}) has been proposed as
an hypothesis which leads to $U_{e3} = 0$ and one massless neutrino. 
As discussed in \cite{mt0}, any neutrino mass matrix which yields
$U_{e3} = 0$ must be invariant under a $Z_2$ symmetry. 
Therefore, $m_\nu$ in Eq.~(\ref{eq:mnu}) must also be invariant under 
some $Z_2$.  This $Z_2$ is easily seen to be the generalized 
$\mu$--$\tau$ symmetry defined in Eq.~(\ref{eq:S}). Indeed if we
identify 
\be
\cos 2 \theta = \frac{c^2 - 1}{1 + c^2} \mbox{ and } 
\sin 2 \theta = \frac{2 c}{1 + c^2} \, ,
\ee
then the mass matrix is invariant according to 
\be \label{eq:mnuS}
S^{-1} \, m_\nu \, S = m_\nu \, .
\ee

While the scaling form for $m_\nu$ satisfies the above equation, it is not the most general
form implied by the invariance under $S$. It is easy to see that the most general $m_\nu$ invariant under   
\be \label{eq:nuLS2} 
\nu_L \rightarrow S \, \nu_L 
\ee
is given by 
\be \label{eq:mnugeneral}
m_\nu = 
\left( 
\bad 
a & b & b/c \\
\cdot & B + C \, \cos 2\theta & C \, \sin 2 \theta\\
\cdot & \cdot & B - C \, \cos 2\theta \\
\ea 
\right) .
\ee
This matrix gives $U_{e3} = 0$ but all eigenvalues are non-zero. 
It reduces to the scaling form in Eq.~(\ref{eq:mnu}) if 
$$ B=C = \frac{d}{c \sin 2\theta}\,,$$ 
in which case one gets the inverted hierarchy structure.

However, scaling follows from the generalized $Z_2$ invariance if 
one additionally assumes that 
$m_\nu$ is obtained from the type-I see-saw mechanism. 
In this case, Eq.~(\ref{eq:nuLS2}) is sufficient to imply scaling. 
In the see-saw mechanism, the mass matrix $m_\nu$ is obtained through 
$m_\nu = -m_D^T \, M_R^{-1} \, m_D$, where $m_D$ is the Dirac mass
matrix and $M_R$ the Majorana mass matrix for the 
right-handed (RH) neutrinos. The Lagrangian reads 
\be
{\cal L} = \frac 12 \, \overline{N_R} \, M_R \, N_R^c + 
\overline{N_R} \, m_D \, \nu_L + h.c.
\ee 
Eq.(\ref{eq:nuLS2}) implies now 
that $m_D \, S= m_D$. This in turn implies 
first that $m_\nu= -m_D^T \, M_R^{-1} \, m_D$ takes 
the generalized $\mu$--$\tau$ invariant form given in
Eq.~(\ref{eq:mnu}). Secondly, it implies an $m_D$
satisfying Eq.~(\ref{eq:md}). 
As a consequence, the resulting $m_\nu$ also possesses the scaling 
form with inverted hierarchy. 
Note that the above results follow for arbitrary non-singular 
$M_R$. Our only assumption is the type-I see-saw and the 
symmetry defined by Eq.~(\ref{eq:nuLS2}). 

In the context of $\mu$--$\tau$ symmetry 
one often \cite{rabi_mutau} applies an additional 
$Z_2$ symmetry which exchanges the 
RH neutrino fields $\nu_{\mu R}\leftrightarrow \nu_{\tau R}$. This  
implies a form of $M_R$ in analogy to Eq.~(\ref{eq:mnumt}) and in
total the mass matrices read for $c=1$
\be \label{eq:ssmt}
m_D = 
\left( 
\bad 
a_1 & d_1 & d_1 \\
a_2 & d_2 & d_3 \\
a_2 & d_3 & d_2 
\ea
\right) \mbox{ and } 
M_R = 
\left( 
\bad
W & X & X \\
\cdot & Y & Z \\ 
\cdot & \cdot & Y 
\ea
\right) .
\ee
Note that this second $Z_2$ 
is neither necessary nor sufficient to obtain a 
$\mu$--$\tau$ symmetric $m_\nu$. The low energy $\mu$--$\tau$ symmetry
follows purely from the left-handed $Z_2$ symmetry as we have seen.

\section{\label{sec:yb}Scaling and Leptogenesis}

Scaling with $c = 1$ leads to the same mixing pattern as  
$\mu$--$\tau$ symmetry. However, their implications at high energy 
can differ.  
Thus both leptogenesis and the lepton flavor violation pattern can be 
different. Leptogenesis in the presence of $\mu$--$\tau$ symmetry was 
considered in \cite{rabi_mutau} and it was shown that the exact 
$\mu$--$\tau$ symmetry implies vanishing lepton asymmetry if there are
only two RH neutrinos. If three RH neutrinos are assumed then the 
lepton asymmetry is related to the solar neutrino mass-squared
difference. 
We will show here that in the case of scaling even for 2 RH neutrinos 
the lepton asymmetry can be non-zero. 
Interestingly, the lepton asymmetry in this simple case 
is related to the solar scale and coincides with the one obtained in 
case of three generations by Mohapatra and Nasri in
Ref.~\cite{rabi_mutau}. 
We trace this coincidence by considering the case 
of three RH neutrinos with scaling and show that one of the 
neutrinos decouples, leading essentially to the two RH neutrino result. 
In addition, we also evaluate various flavor lepton asymmetries 
in these cases.

\subsection{\label{sec:2NR}Scaling with two heavy Neutrinos and 
Leptogenesis}

With only two RH neutrinos present, the Dirac and Majorana mass matrices 
can be written as 
\be
m_D = \left( 
\bad 
A_1 & B & B/c \\
A_2 & D & D/c
\ea  
\right) \mbox{ and } 
M_R = V_R^\ast \, D_R \, V_R^\dagger  \, , 
\ee
Here, we have imposed the scaling form of $m_D$ while 
$M_R$ is kept general.  Thus $V_R$ is a general unitary $2\times2$ 
matrix and  $D_R = {\rm diag}(M_1, \, M_2)$ with $M_2 > M_1$. 
The $3\times3$ mass matrix 
of the light neutrinos reads 
\be \label{eq:temp}
m_\nu = - m_D^T \, V_R \, D_R^{-1} \, V_R^T \, m_D \equiv 
-\tilde{m}_D^T \, D_R^{-1} \, \tilde{m}_D \, .
\ee
We have defined here $\tilde{m}_D = V_R^T \, m_D$. 
The most general matrix diagonalizing $m_\nu$ will be called $U$, 
and is defined by  
\be \label{eq:U2}
U^T \, m_\nu \, U = D_\nu = {\rm diag}(m_1 \, e^{i \alpha_1} , 
\, m_2 \, e^{i \alpha_2} , \, 0) \, ,
\mbox{ where } U = R_{23} \, U_{12}\,. 
\ee
We have kept the phases of the eigenvalues of $m_\nu$. 
Here $R_{23}$ is a rotation in 23-space
\be
R_{23} = \left( \bad 
1 & 0 & 0 \\
0 & c_{23} & -s_{23} \\
0 & s_{23} & c_{23} 
\ea
\right) \mbox{ with } 
c_{23} = \frac{c}{\sqrt{1 + c^2}}\,,~
s_{23} = \frac{1}{\sqrt{1 + c^2}}
\ee
and 
\be
U_{12} = 
\left( 
\baz 
u & 0 \\
0 & 1 
\ea
\right) \mbox{ with } 
u = P(\alpha) 
\left( 
\baz 
c_{12} & s_{12} \\
-s_{12} & c_{12} 
\ea 
\right) P(\beta) \, .
\ee
The phase matrices are defined via 
$P(\gamma) = {\rm diag}(e^{i \gamma},\,e^{-i\gamma})$.  
Note that $\theta_{12}$ is the solar neutrino mixing angle and 
$\theta_{23}$ the atmospheric neutrino mixing angle with 
$\tan^2 \theta_{23} = 1/c^2$. 

Knowing the matrix $U$ allows us to go into the neutrino mass basis: 
\be \label{eq:massbas}
D_\nu = - U_{12}^T \, R_{23}^T \, \tilde{m}_D^T \, D_{R}^{-1} \, 
\tilde{m}_D \, R_{23} \, U_{12} \, \equiv 
-{\cal Z}^T \, D_{R}^{-1} \, {\cal Z}\,,
\ee
where ${\cal Z}$ is a $3 \times 2$ matrix defined by 
\be \label{eq:Z2}
{\cal Z} = \tilde{m}_D \, R_{23} \, U_{12} = 
\left( \bad
Z_{11} & Z_{12} & 0 \\
Z_{21} & Z_{22} & 0 
\ea
\right) \mbox{ and } 
Z_{ij} = (V_R^T \, m_D' \, u)_{ij} \, .
\ee
The definition of $m_D'$ is 
\be \label{eq:mds} 
m_D' = \left( \baz 
A_1 & B/c_{23} \\
A_2 & D/c_{23} 
\ea \right) .
\ee 
With Eq.~(\ref{eq:massbas}) it follows 
\be
Z^T \, D_R^{-1} \, Z = - 
\left( 
\baz 
m_1 \, e^{i \alpha_1} & 0 \\
0 & m_2 \, e^{i \alpha_2} 
\ea
\right) .
\ee
From this relation one obtains constraints on the $Z_{ij}$: 
\bea \label{eq:Z_rel} \D 
m_1 \, e^{i \alpha_1} = -\frac{Z_{11}^2}{M_1} \left( 
1 + \frac{Z_{21}^2}{Z_{11}^2} \frac{M_1}{M_2}
\right) \equiv 
-\frac{Z_{11}^2}{M_1} \left(1 + r \, e^{i\rho} \right) , \\ \D 
m_2 \, e^{i \alpha_2} = -\frac{Z_{22}^2}{M_2} \left( 
1 + \frac{Z_{21}^2}{Z_{11}^2} \frac{M_1}{M_2}
\right) \equiv 
-\frac{Z_{22}^2}{M_2} \left(1 + r \, e^{i\rho} \right) ,\\  \D 
\frac{Z_{12}}{Z_{22}} = - \frac{Z_{21}}{Z_{11}} \frac{M_1}{M_2} 
= -\sqrt{r} \, \sqrt{\frac{M_1}{M_2}} \, e^{i \rho/2} \, .
\eea
Since in the inverted hierarchy with $m_3 = 0$ it holds that 
$m_2 \simeq m_1$, we see that we need to fulfill the requirement 
$|Z_{11}^2 /M_1| \simeq |Z_{22}^2 /M_2|$. 
The above equations imply 
\be \label{eq:dms}
\dms = |1 + r \, e^{i\rho}|^2 \left( 
\frac{|Z_{22}|^4}{M_2^2} - \frac{|Z_{11}|^4}{M_1^2}
\right) .
\ee
The decay asymmetries of the heavy neutrinos are determined 
in the flavor basis and in terms of 
$\tilde{m}_D$ defined in Eq.~(\ref{eq:temp}). We can write  
\be
\tilde{m}_D = {\cal Z} \, U_{12}^\dagger \, R_{23}^T \, ,
\ee
from which it follows that 
\be
\tilde{m}_D\tilde{m}_D^\dagger = {\cal Z} {\cal Z}^\dagger 
= \left( 
\baz 
Z Z^\dagger & 0 \\
0 & 0 
\ea
\right) 
\, .
\ee
With the relations for the $Z_{ij}$ in Eq.~(\ref{eq:Z_rel}) it follows 
for the effective mass governing the wash-out: 
\be \label{eq:m1t}
\tilde{m}_1 = \frac{(Z Z^\dagger)_{11}}{M_1} = 
\frac{m_1 + r \, m_2}{|1 + r \, e^{i\rho}|} \, .
\ee
Since $m_3 = 0$ and an inverted hierarchy is present, 
we have, $m_2 \simeq m_1 \simeq \sqrt{\dma}$. Consequently, 
$\tilde{m}_1$ is of order $\sqrt{\dma}$ (this value is in fact the
limit of $\tilde{m}_1$ for both $r \ra 0$ and $r \ra \infty$). 
We are thus in the ``strong wash-out'' regime, in which case 
there is little dependence on the initial conditions, and 
the efficiency factor is of order $10^{-2}$ or $10^{-3}$. 
More importantly, we find  with Eq.~(\ref{eq:dms}) 
\be
{\rm Im} \left\{ (Z Z^\dagger)_{12}^2 \right\}  = M_1 \, M_2 \, r \, 
\frac{\dms}{|1 + r \, e^{i\rho}|^2} \, \sin \rho \, ,
\ee
and with the natural and usual assumption $M_2 \gg M_1$ the 
decay asymmetry (see the Appendix for the relevant formulae) 
for unflavored leptogenesis reads 
\be \label{eq:eps1}
\varepsilon_1 = - \frac{3}{16 \pi \, v^2} \frac{M_1}{\tilde{m}_1} \, r \,  
\frac{\dms}{|1 + r \, e^{i\rho}|^2} \, \sin \rho \, .
\ee
Numerically, with $\tilde{m}_1 \simeq \sqrt{\dma}$, we have 
\be
\varepsilon_1 \simeq  -3.1 \times 10^{-6} 
\left(\frac{M_1}{10^{12}~\rm GeV} \right) 
\, \frac{r}{|1 + r \, e^{i\rho}|^2} \, \sin \rho \, ,
\ee
where we have inserted the current best-fit values 
\cite{bari} for $\dms = 7.67 \times 10^{-5}$ eV$^2$ and for 
$\dma = 2.39 \times 10^{-3}$ eV$^2$. 
With an efficiency factor $\eta$ of order $10^{-2}$ or $10^{-3}$ 
the baryon asymmetry of order $Y_B \simeq 10^{-2} \, \varepsilon_1 
\, \eta \stackrel{!}{\simeq} 10^{-10}$ can easily be generated.\\ 

Unlike in the $\mu$--$\tau$ symmetric case, one obtains a non-zero 
asymmetry with two RH neutrinos independent of the value of $c$. 
More interestingly, the lepton asymmetry obtained in 
Eq.~(\ref{eq:eps1}) coincides with 
Eq.~(20) in the first paper of \cite{rabi_mutau}. 
Note that that analysis corresponds to a quite different situation 
than the one considered here namely, three RH neutrinos and 
$\mu$--$\tau$ symmetry for $m_D$ and $M_R$.  
We trace the origin of this result by 
considering a full three generation case in the next Subsection. 
Prior to this we discuss leptogenesis flavor effects
\cite{flav} which can be important  for $M_1 \leq 10^{12}$ GeV. 
These effects were not considered in \cite{rabi_mutau}. We show in the
Appendix that individual flavor asymmetries vanish if 
$\mu$--$\tau$ symmetry is exact and only two RH neutrinos are present. 
This is not the case with scaling. 
The individual flavored decay asymmetries for our case read 
\be
\varepsilon_1^\mu = c_{23}^2 \, (\varepsilon_1 - \varepsilon _1^e)\,,~
\varepsilon_1^\tau = s_{23}^2 \, (\varepsilon_1 - \varepsilon _1^e)
\ee
and 
\begin{eqnarray}
\nonumber 
 \varepsilon_1^e &=& - \frac{3 \, M_1}{16 \pi \, v^2 \, \tilde{m}_1 \, 
|1 + r \, e^{i\rho}|^2} \left( 
r \left(m_2^2 \, s_{12}^2 - m_1^2 \, c_{12}^2  \right) \sin \rho + \right. \\
& & \left.   
c_{12} \, s_{12} \, \sqrt{m_1 \, m_2 \, r} \left( 
(m_1 - m_2 \, r) \, \sin(\alpha_1 - \alpha_2  -4\beta-\rho)/2 + 
\right. \right. \\ \nonumber 
& & \left. \left.   
(m_1 \, r - m_2 ) \, \sin(\alpha_1 - \alpha_2  -4\beta + \rho)/2 
\right)
\right)  .
\end{eqnarray}
The individual wash-out parameters are 
\begin{eqnarray} \label{eq:washput}
\tilde{m}_1^e & = & \frac{1}{|1 + r \, e^{i\rho}|} 
\left( m_1 \, c_{12}^2 + r \, m_2 \, s_{12}^2 - 2 \, c_{12} \, 
s_{12} \, \sqrt{m_1 \, m_2 \, r} \, 
\cos (\alpha_1-\alpha_2-4\beta-\rho)/2 \right) \, , \nonumber \\
\tilde{m}_1^\mu & = & 
\frac{c_{23}^2}{|1 + r \, e^{i\rho}|} \left( r \, m_2 \, c_{12}^2 + 
m_1 \, s_{12}^2 + 2 \, c_{12} \, s_{12}\sqrt{m_1 \, m_2 \, r} \, 
\cos (\alpha_1-\alpha_2-4\beta-\rho)/2 \right) \, , \nonumber \\
\tilde{m}_1^\tau & = & \tan^2\theta_{23} \, \tilde{m}_1^\mu \, . 
\end{eqnarray}
One can explicitly check that the sum over the individual 
decay asymmetries and wash-out parameters results 
in the unflavored $\varepsilon_1$ and 
$\tilde{m}_1$ from Eqs.~(\ref{eq:eps1}) and (\ref{eq:m1t}),
respectively. 
Note that the individual flavor symmetries are generically
proportional to the atmospheric mass scale although the combined sum 
depends on the solar scale.

Consider the limit $r = 1$. The term 
proportional to $\sin \rho$ dominates in $\varepsilon_1^e$, and 
\be
\varepsilon_1 - \varepsilon_1^e \stackrel{r \ra 1 }{\simeq}
- \frac{3}{16 \pi \, v^2} \frac{M_1}{\tilde{m}_1} \, 
\frac{\dma \, \cos 2 \theta_{12}}{|1 + e^{i\rho}|^2} \, \sin \rho \, .
\ee
We have used here again 
the fact that $m_2 \simeq m_1 \simeq \sqrt{\dma}$. Note that 
$\dma \, \cos 2 \theta_{12}$ is for the inverted hierarchy 
the minimal value of the effective
mass governing neutrino-less double beta decay. 

The final baryon asymmetry is given by \cite{flav}
\be \label{asymmetry}
Y_B \simeq \left\{
\baz
-0.01 \, \varepsilon_1 \, \eta(\tilde{m}_1)
& \mbox{one-flavor}\, ,\\[0.2cm]
-0.003 %\frac{12}{37 \, g^\ast}
\left(
(\varepsilon_1^e + \varepsilon_1^\mu) \, \eta\left(\frac{417}{589}
(\tilde{m}_1^e + \tilde{m}_1^\mu) \right) +
\varepsilon_1^\tau \, \eta \left(\frac{390}{589}
\tilde{m}_1^\tau \right) \right)
& \mbox{two-flavor}\, ,\\[0.2cm]
-0.003 %-\frac{12}{37 \, g^\ast}
\left(
\varepsilon_1^e \, \eta \left(\frac{151}{179}
\, \tilde{m}_1^e  \right) +
\varepsilon_1^\mu \, \eta \left(\frac{344}{537} \, \tilde{m}_1^\mu \right) +
\varepsilon_1^\tau \, \eta \left(\frac{344}{537} \,
\tilde{m}_1^\tau \right) \right) & \mbox{three-flavor}\,.
\ea \right.
\ee
Here we have given separate expressions
for one-, two- and three-flavored
leptogenesis \cite{flav}. The three-flavor case occurs for
$M_1 \ls 10^9$ GeV and corresponds to the situation when both 
$\mu$ and $\tau$ Yukawa coupling induced processes dominate 
over the Dirac neutrino couplings. The one-flavor case occurs 
$M_1 \gs 10^{12}$ GeV,
and the two-flavor case (with the tau-flavor decoupling first
and the sum of electron- and muon-flavors, which act indistinguishably)
applies in between. In case of the MSSM we need to multiply these 
mass values with $1 + \tan^2 \beta$. The efficiency $\eta$ is a 
function of the wash-out parameter $\tilde m$ and its 
form is given in the Appendix. 
Let us give one example to show that the correct baryon asymmetry can
be generated: 
it follows from Eq.~(\ref{eq:washput}) that there can be range of 
parameters for which $\tilde{m}_1^\mu \simeq \tilde{m}_1^\tau$ 
dominates over $\tilde{m}_1^e$ or vice versa. 
If $\tilde{m}_1^e$ is suppressed compared to the other two 
then one finds from Eq.~(\ref{eq:washput}) $\tilde{m}_1^\mu \simeq 
\tilde{m}_1^\tau \sim \frac{m_1+rm_2}{|1+r e^{i \rho}|}$. 
In this case, Eq.~(\ref{asymmetry}) leads to $$Y_B\simeq 10^{-10}\,,$$ 
where we have considered the three flavor case  
and chosen $M_1 = 10^9$ GeV, $r=1$ and $\rho=\frac{\pi}{4}$. 
It follows that the scaling case considered here 
can generate the required baryon asymmetry.\\

\subsection{\label{sec:3NR}Scaling with 3 right-handed Neutrinos and 
$\mu$--$\tau$ symmetry}

An effective 
two heavy neutrino framework discussed in the previous 
Subsection can be obtained quite easily. 
We have seen that in the framework of the see-saw mechanism 
invariance under the transformation $\nu_L \rightarrow S \, \nu_L$, 
where $S$ is the $Z_2$ 
generator defined in Eq.~(\ref{eq:S}), leads to a low energy 
mass matrix obeying scaling. Consider now in addition to 
this $Z_{2L}$ symmetry an additional $Z_{2R}$ under 
which the heavy neutrinos are invariant: 
\be \label{eq:SNR}
N_R \rightarrow S \, N_R \, .
\ee
This invariance implies 
\be \label{eq:implies}
S \, m_D = m_D \mbox{ and } S \, M_R \, S = M_R\, .
\ee
Together with the $Z_{2L}$ symmetry from above we have in total a  
$Z_{2L} \times Z_{2R}$ symmetry.  
The Dirac mass matrix $m_D$ which satisfies simultaneously 
$m_D \, S = m_D $ and $S \, m_D = m_D$ can be written as 
\be
m_D = \left( 
\bad
A_1 & B & B \, s_{23}/c_{23} \\
A_2 \, c_{23} & D \, c_{23} & D \, s_{23} \\
A_2 \, s_{23} & D \, s_{23} & D \, s_{23}^2/c_{23}
\ea
\right) .
\ee
%Note that this form is not obtained from 
%$S \, m_D \, S = m_D$. First we need invariance 
%under $m_D \, S = m_D $ and then under 
%$S \, m_D \, S$ (or vice versa). 
The most general solution of the $Z_{2R}$-invariance 
of the heavy neutrinos in 
Eq.~(\ref{eq:implies}) is (see also \cite{mt0})
\be
M_R = \left( 
\bad
A & B & B/c \\
\cdot & F \, (c - 1/c) + G & F \\
\cdot & \cdot & G 
\ea
\right) . 
\ee
The eigenvector corresponding to the eigenvalue $G - F/c$ of this matrix is 
the ``scaling vector'' $(0,-1,c)^T$. The matrix $M_R$ is diagonalized as 
\be
\tilde{V}_R^T \, R_{23}^T \, M_R \, R_{23} \, \tilde{V}_R = D_R 
\mbox{ with }
\tilde{V}_R = \left( \baz 
V_R & 0 \\
0 & e^{-i \phi_3/2}
\ea \right),
\ee
where $V_R$ is a general unitary $2\times2$ matrix and $\phi_3$ the phase 
of the third eigenvalue of $M_R$. The light neutrino mass matrix 
can be written as 
\be
m_\nu = - m_D^T \, R_{23} \, \tilde{V}_R \, D_R^{-1} \, \tilde{V}_R^T \, 
R_{23}^T \, m_D \equiv 
\tilde{m}_D^T \, D_R^{-1} \, \tilde{m}_D \, .
\ee
Obviously, $m_\nu$ is diagonalized by the matrix $U$ from Eq.~(\ref{eq:U2}). 
Consequently, in the neutrino mass basis we have 
\be
D_\nu = - {\cal \hat{Z}}^T \, D_R^{-1} \, {\cal \hat{Z}} \, ,
\ee
where 
\be \label{eq:Z3}
{\cal \hat{Z}} \equiv \tilde{m}_D \, R_{23} \, U_{12} 
= V_R^T \, R_{23}^T \, m_D \, R_{23} \, U_{12} 
= \left( \baz 
Z & 0 \\ 
0 & 0 
\ea
\right),  
\ee
and we have used the relation 
\be
R_{23}^T \, m_D \, R_{23} = \left( \baz 
m_D' & 0 \\
0 & 0
\ea
\right)
\ee
with the definition of $m_D'$ in Eq.~(\ref{eq:mds}). Note that the 
matrix $Z$ appearing in Eq.~(\ref{eq:Z3}) exactly coincides with the $Z$ 
defined in Eq.~(\ref{eq:Z2}) in case of two heavy right-handed neutrinos. 
Likewise $\tilde{m}_D  \tilde{m}_D^\dagger$ also coincides with the 
two right-handed neutrino case. Thus the three right-handed neutrino case 
with $Z_{2L} \times Z_{2R}$ symmetry defined in this Subsection gives the 
same decay and baryon asymmetry and neutrino masses as the two 
right-handed neutrino case treated in Section \ref{sec:2NR}. 

The structures of the $Z_{2L}\times Z_{2R}$ symmetric $m_D$ and $M_R$ 
reduce to the $\mu$--$\tau$ symmetric structures when $c = 1$. 
The decoupling obtained here is the same as obtained in the case 
of 3 RH neutrinos and $\mu$--$\tau$ symmetric see-saw 
considered in \cite{rabi_mutau}, except that $c$ needs not to be 1 here. 
Because of the specific structures of the mass matrices, one combination of 
$\nu_{\mu R}$ and $\nu_{\tau R}$ decouples. The matrix 
$M_R$ for the remaining two neutrinos does not obey 
$\mu$--$\tau$ symmetry. Thus the decoupling limit of this
$Z_{2L}\times Z_{2R}$ invariant case corresponds to 
a scaling like situation with only two RH neutrinos for 
arbitrary $M_R$, and the lepton asymmetries in these two cases
coincide.

\subsection{Scaling and Lepton Flavor Violation} 
Another place in which the predictions of scaling and 
$\mu$--$\tau$ symmetry can differ is lepton flavor violation 
in supersymmetric see-saw models. 
In such supersymmetric type I see-saw frameworks with universal (mSUGRA) 
boundary conditions at the GUT scale $M_X$, the branching ratios 
for lepton flavor violating (LFV) charged lepton decays 
$\ell_i \ra \ell_j \, \gamma$ are proportional to \cite{lfv}
\be
\mbox{BR}(\ell_i \ra \ell_j \, \gamma) \propto 
\mbox{BR}(\ell_i \ra \ell_j \, \nu \overline{\nu}) \, 
\left|(\tilde{m}_D^\dagger \, L \, \tilde{m}_D)_{ij} \right|^2  ,
\ee
where $L_{ij} = \delta_{ij} \, 
\log M_i/M_X$ and $M_i$ are the individual heavy neutrino
masses. Here we have rotated $m_D$ to $\tilde{m}_D = V_R^T \, m_D$, 
where $M_R = V_R^\ast \, M_R^{\rm diag} \, V_R^\dagger$. 
One can check that as long as $m_D$ is given by 
Eq.~(\ref{eq:md}) the relation 
\be \label{eq:lfv1}
\frac{\left|(\tilde{m}_D^\dagger \, L \, \tilde{m}_D)_{12} \right|^2}
{\left|(\tilde{m}_D^\dagger \, L \, \tilde{m}_D)_{13} \right|^2} 
= c^2 = \cot^2 \theta_{23} \, 
\ee
holds for arbitrary $M_R$. Hence, the branching ratios of 
$\mu \ra e \gamma$ and $\tau \ra e \gamma$ are close to each other 
(up to a normalization factor BR($\tau \ra e \, 
\nu \overline {\nu}) \simeq 0.178$). Note that this implies that 
$\tau \ra e \gamma$ will be too rare to be observable. 
We can compare this with 
$\mu$--$\tau$ symmetric see-saw, i.e., the matrices in
Eq.~(\ref{eq:ssmt}) with which it follows:  
\be \label{eq:lfvL3}
\frac{\left|(\tilde{m}_D^\dagger \,  \tilde{m}_D)_{12} \right|^2}
{\left|(\tilde{m}_D^\dagger \, \tilde{m}_D)_{13} \right|^2} 
= 1 \, . 
\ee
Note that we have not included here the diagonal matrix $L$, i.e., 
the unity of this ratio holds only up to potentially significant
logarithmic corrections \cite{WR}. This is in contrast to see-saw and 
scaling, cf.\ Eq.~(\ref{eq:lfv1}).

\section{\label{sec:concl}Conclusions and Summary}

Maximal atmospheric mixing and vanishing $U_{e3}=0$ are two important
predictions generally attributed to $\mu$--$\tau$ symmetry. 
The scaling Ansatz proposed in Ref.~\cite{scal1,scal2} is another 
possibility which leads to $U_{e3} = 0$ and potentially also to 
maximal $\theta_{23}$. In a 
see-saw framework, the usually applied $\mu$--$\tau$ symmetry 
heavily restricts the form of the mass matrices $m_D$ and $M_R$, 
whereas the scaling Ansatz does not. 
This paper was devoted to understand relationships between 
scaling and $\mu$--$\tau$ symmetry and to discuss some phenomenological 
implications of scaling. In particular, we have shown here that: 

\begin{itemize}
\item the scaling hypothesis is a special case of the 
generalized $\mu$--$\tau$ symmetry leading to $U_{e3}=0$
and an inverted hierarchy; 

\item the only way to derive scaling in the type-I see-saw mechanism 
is to have a Dirac neutrino mass matrix satisfying Eq.~(\ref{eq:md}) 
as long as two neutrinos are required to have non-zero masses; 

\item scaling follows from a generalized $\mu$--$\tau$ symmetry
in the type-I see-saw model; 

\item the scaling hypothesis in the limit $c = 1$ gives the same
lepton mixing as $\mu$--$\tau$ symmetry, but the predictions for
leptogenesis and lepton flavor violation differ. 
Unlike the $\mu$--$\tau$ symmetry, a simple model with scaling 
and two RH neutrino leads to non-vanishing lepton asymmetry, 
which is proportional to the solar neutrino mass-squared difference; 
 
\item in the simple example with scaling and two RH neutrinos we 
have shown that it is possible to obtain the required baryon 
asymmetry through both flavored and unflavored leptogenesis.  
The same should be true in a more general example with three 
RH neutrinos which contains more parameters; 

\item the model with two RH neutrinos and $m_D$ in the scaling 
form may be regarded as a limit of the 3 RH neutrino model with 
$\mu$--$\tau$ symmetry.  We have explicitly shown that the 
latter model reduces to the former after decoupling and 
both lead to the same lepton asymmetry.

\end{itemize}

The scaling hypothesis and $\mu$--$\tau$ symmetry are thus related, 
though some interesting differences exist. Finally, 
it is worth mentioning that the scaling hypothesis is a highly 
attractive possibility to generate an inverted neutrino mass
hierarchy. In contrast to the usually considered $L_e - L_\mu -
L_\tau$ flavor symmetry, no breaking is required in order to reach
agreement with data. The many interesting features of scaling and/or 
the inverted hierarchy in neutrino oscillations or neutrino-less 
double beta decay are an easy and soon to be performed test 
of this Ansatz.

\vspace{0.3cm}
\begin{center}
{\bf Acknowledgments}
\end{center}
W.R.~wishes to thank the Physical Research Laboratory in Ahmedabad  
and the organizers of the workshop 
``Astroparticle Physics - A Pathfinder to New Physics'' in Stockholm, 
where part of this paper was written, for hospitality. 
This work was supported by the ERC under the Starting Grant 
MANITOP and by the Deutsche Forschungsgemeinschaft 
in the Transregio 27 ``Neutrinos and beyond -- weakly interacting 
particles in physics, astrophysics and cosmology'' (W.R.). 

\renewcommand{\theequation}{A\arabic{equation}}
\setcounter{equation}{0}
\renewcommand{\thetable}{A\arabic{table}}
\setcounter{table}{0}

\begin{appendix}
\section{\label{sec:A}$\mu$--$\tau$ symmetric See-Saw and Leptogenesis}

An interesting point of $\mu$--$\tau$ symmetric 
see-saw scenarios is that in case of two heavy neutrinos there is no
successful leptogenesis \cite{rabi_mutau}. We repeat here for completeness 
the derivation of this result and add 
its invariance under the presence or absence of 
flavor effects: in a $\mu$--$\tau$ symmetric 
see-saw framework with two heavy neutrinos the mass matrices read 
(writing here explicitly all phases) 
\be
m_D = 
\left(
\bad 
a \, e^{i \alpha} & b \, e^{i \beta} & d \, e^{i \delta} \\
a \, e^{i \alpha} & d \, e^{i \delta} & b \, e^{i \beta} 
\ea
\right) \mbox{ and } 
M_R = 
\left(
\baz 
M_{11} \, e^{i \phi_{11}} & M_{12} \, e^{i \phi_{12}} \\
M_{12} \, e^{i \phi_{12}} & M_{11} \, e^{i \phi_{11}}
\ea
\right) .
\ee
Diagonalizing $M_R$ via 
$M_R = V_R^\ast \, M_R^{\rm diag} \, V_R^\dagger$ gives 
\bea
V_R = \left( 
\baz 
-\frac{1}{\sqrt{2}} & \frac{1}{\sqrt{2}} \\
\frac{1}{\sqrt{2}} & \frac{1}{\sqrt{2}}
\ea
\right) 
\left( 
\baz 
e^{-i \phi_1} & 0 \\
0 & e^{-i \phi_2}
\ea
\right) \\
\mbox{ and } 
M_R^{\rm diag} = 
\left(
\baz
M_1 & 0 \\
0 & M_2 
\ea
\right) = 
\left(
\baz
|M_{11} \, e^{i \phi_{11}} - M_{12} \, e^{i \phi_{12}}| & 0 \\
0 & |M_{11} \, e^{i \phi_{11}} + M_{12} \, e^{i \phi_{12}}| 
\ea
\right) ,
\eea
where $\phi_{1,2}$ are the phases of $M_1$ and $M_2$. The decay
asymmetries of the heavy neutrinos, to which the baryon asymmetry 
is proportional are in general \cite{lepto}
\bea \label{eq:eps}
\epsilon_1^\alpha = 
\frac{\D 1}{\D 8 \pi \, v^2} 
\, \frac{\D  M_1}{\D \tilde{m}_1} 
\sum \limits_j \left[ 
 {\rm Im} \left\{ (\tilde{m}_D)_{1 \alpha} \, 
(\tilde{m}_D^\dagger)_{\alpha j} 
\, (\tilde{m}_D \, \tilde{m}_D^\dagger)_{1j} \right\}
\, f(x_j) \right. \\
\left. 
+ 
{\rm Im} \left\{ (\tilde{m}_D)_{1 \alpha} \, 
(\tilde{m}_D^\dagger)_{\alpha j} 
\, (\tilde{m}_D \, \tilde{m}_D^\dagger)_{j1} \right\}
\, g(x_j)
\right] ,
\eea
where the wash-out is governed by 
\be
\tilde{m}_1 = 
\frac{(\tilde{m}_D \, \tilde{m}_D^\dagger)_{11}}{M_1} \, ,
\ee 
which is a sum over the individual wash-out parameters 
\be
\tilde{m}_1^\alpha = 
\frac{(\tilde{m}_D)_{1 \alpha} \, (\tilde{m}_D^\dagger)_{\alpha
1}}{M_1} \, .
\ee
The wash-out parameters need to be inserted in the approximate formula
\be \label{washout}
\eta(x) \simeq
\left(
\frac{8.25 \times 10^{-3}~{\rm eV}}{x} +
\left(\frac{x}{2 \times 10^{-4}~{\rm eV}}\right)^{1.16}
\right)^{-1}\,.
\ee
The functions $f$ and $g$ depend on   
$x_j = M_j^2 /M_1^2$ and are given by 
\be \label{eq:fij}
f(x) = \sqrt{x} \, \left( 
\frac{2 - x}{1 - x} - (1 + x) \, \ln  \left(1 + \frac 1x \right) 
\right) \simeq - \frac{3}{2 \sqrt{x}} \, ,
\ee
where the approximate expression holds for $x \gg 1$ and 
\be \label{eq:gij}
g(x) = \frac{1}{1 - x} \, .
\ee
Note that the terms proportional to $g(x_j)$ drop out when summed 
over $\alpha$ (i.e., when flavor effects play no role) and that they
are suppressed by a factor of $M_1/M_j$ 
with respect to the terms proportional to $f(x_j)$. 
With 
$m_D$ given as above and $\tilde{m}_D = V_R^T \, m_D$ it follows 
\be
\tilde{m}_D \, \tilde{m}_D^\dagger = 
\left(
\baz 
b^2 + d^2 - 2 \, b \, d \, \cos (\beta - \delta) & 0\\
0 & 2 \, a^2 + b^2 + d^2 + 2 \, b \, d \, \cos (\beta - \delta) 
\ea
\right) 
\ee
and consequently $\epsilon_1^e = \epsilon_1^\mu = \epsilon_1^\tau =
0$. Hence, the baryon asymmetry is non-zero in the case of 
a $\mu$--$\tau$ symmetric two heavy neutrino framework \cite{rabi_mutau}. 
We add here to the result from Ref.~\cite{rabi_mutau} that 
flavor effects do not change the situation. We furthermore note 
that regarding LFV in SUSY see-saw scenarios the relation 
\be
\frac{\left|(\tilde{m}_D^\dagger \, L \, \tilde{m}_D)_{12} \right|^2}
{\left|(\tilde{m}_D^\dagger \, L \, \tilde{m}_D)_{13} \right|^2} 
= 1 \, 
\ee
holds. 
The branching ratios of $\mu \ra e \gamma$ and $\tau \ra e \gamma$ are 
identical (up to a normalization factor 
BR($\tau \ra e \, \nu \overline {\nu}) \simeq 0.178$) 
even when the logarithmic factors in $L$ are taken into account, 
cf.~Eq.~(\ref{eq:lfvL3}). Note that this implies that 
$\tau \ra e \gamma$ will be too rare to be observable. 
For completeness, we find for the 
expression relevant to $\tau \ra \mu \gamma$ that 
$\left|(\tilde{m}_D^\dagger \, L \, \tilde{m}_D)_{12} \right|^2 = 
a^2 \, L_2^2 \, (b^2 + d^2 + 2 \, b \, d 
\, \cos (\beta - \delta) )$.

\end{appendix}

\end{document}